\documentclass[12pt,a4paper]{article}

\def\d{\partial}
\def\defpar{\vartheta}

\def\g{h}

\def\beann{\begin{eqnarray*}}
\def\eeann{\end{eqnarray*}}
\def\beq{\begin{equation}}
\def\eeq{\end{equation}}
\def\ba{\begin{array}}
\def\ea{\end{array}}
\def\ben{\begin{enumerate}}
\def\een{\end{enumerate}}
\def\bea{\begin{eqnarray}}
\def\eea{\end{eqnarray}}
\def\beann{\begin{eqnarray*}}
\def\eeann{\end{eqnarray*}}
\def\beq{\begin{equation}}
\def\eeq{\end{equation}}
\def\ba{\begin{array}}
\def\ea{\end{array}}
\def\ben{\begin{enumerate}}
\def\een{\end{enumerate}}

\def\5{\bar }
\def\6{\partial }
\def\7{\hat }
\def\4{\tilde }

\newcommand{\qcommut}[2]{{[#1\stackrel{*}{,}#2]}}

\newcommand{\tr}{{\rm \,Tr}\,}
\def\half{{\frac{1}{2}}}

\def\beq{\begin{equation}}

\def\eeq{\end{equation}}

\begin{document}

\begin{titlepage}

\begin{flushright}
ULB-TH/01-18\\
hep-th/0201139
\end{flushright}

\begin{centering}

\vspace{0.5cm}

{\bf{\Large Seiberg-Witten maps in the context of the antifield formalism}} \\

\vspace{1.5cm}

{\large Glenn Barnich$^{*}$}

\vspace{.5cm}

Physique Th\'eorique et Math\'ematique,\\ Universit\'e Libre de
Bruxelles,\\
Campus Plaine C.P. 231, B--1050 Bruxelles, Belgium

\vspace{1cm}

{\large Friedemann Brandt}

\vspace{.5cm}

Max-Planck-Institute for Mathematics in the Sciences,\\
Inselstra\ss e 22--26, D--04103 Leipzig, Germany

\vspace{1cm}

{\large Maxim Grigoriev   }

\vspace{.5cm}

Lebedev Physics Institute,
53 Leninisky Prospect,
Moscow 117924, Russia

\vspace{.5cm}

\end{centering}
\vspace{.5cm}

\begin{abstract}
The formulation of Seiberg-Witten maps
from the point of view of consistent
deformations of gauge theories in the context of the Batalin-Vilkovisky
antifield formalism is reviewed. Some additional remarks on noncommutative
Yang-Mills theory are made.
\end{abstract}

\vfill

\footnotesize{$^*$Research Associate of the Belgium National Fund for
Scientific Research.}

\end{titlepage}

\section{Introduction}

Using arguments from string theory, noncommutative Yang-Mills theory
has been shown \cite{Seiberg:1999vs} to be equivalent, by a
redefinition of the gauge potentials and the parameters of the gauge
transformation, to a Yang-Mills theory with standard gauge
symmetries and with an effective action containing, besides the usual
Yang-Mills term, higher dimensional gauge invariant operators.

By considering an expansion in the parameter of noncommutativity
$\vartheta$,
noncommutative Yang-Mills theory can be understood as a consistent
deformation of standard Yang-Mills theory in the sense that the action
and gauge transformations are deformed simultaneoulsy in such a way
that the deformed action is invariant under the deformed gauge
transformations. An appropriate framework for the analysis of such consistent
deformations of gauge theories has been shown \cite{Barnich:1993vg} to be the
antifield-antibracket formalism (see
\cite{Becchi:1975nq,Tyutin:1975qk,Zinn-Justin:1974mc,
Dixon:1975si} in the Yang-Mills context,
\cite{BV} for the generic case and
\cite{Henneaux:1992ig,Gomis:1995he}
for reviews).

By reformulating the question of existence of Seiberg-Witten maps
in this context \cite{Gomis:2000sp,Brandt:2001aa,Barnich:2001mc},
the whole power of the theory of general (anti-)canonical
transformations is available and Seiberg-Witten maps appear as "time-dependent"
canonical transformations that map the gauge structure of the noncommutative
theory to that of the commutative one. In the generic case, this leads to an
appropriate ``open'' version of the gauge equivalence condition, valid only up
to terms vanishing when the equations of motions hold. These
features have been shown \cite{Barnich:2001mc} to be crucial for the
construction of a Seiberg-Witten map in the case of the noncommutative
Freedman-Townsend model.

\section{Equivalent formulations of Seiberg-Witten maps}

Notations
and conventions are as in \cite{Barnich:2001mc}, except that for
later convenience,  we denote the deformation parameter by $\vartheta$ instead
of $g$. Consider a gauge theory determined by the minimal proper
solution $\hat S[\hat\phi,\hat \phi^*;\defpar]$ of the master equation,
\bea
\half(\hat S,\hat S)=0 ,\label{meq}
\eea
and suppose that $\hat S[\hat\phi,\hat \phi^*;\defpar]$ admits
formal power series expansion in the deformation parameter
$\defpar$: $
\hat S[\hat\phi,\hat \phi^*;\defpar]
=\sum_{s=0}^\infty \defpar^s S^{(s)}[\hat\phi,\hat\phi^*]\,.
$
Then, the deformed theory defined by $\hat S$ is a
consistent deformation of the undeformed theory determined by the master
action
\bea
S^{(0)}[\hat\phi,\hat\phi^*]
=
\hat S[\hat\phi,\hat \phi^*;\defpar]\Big|_{\defpar=0},\
\ \half(S^{(0)},S^{(0)})=0\,.
\eea
In what follows
we also use expansions in antifield number; the antifield number of a local
function\footnote{In this context, a local function is a formal power series in
$\defpar$ each term of which depends on the fields, the antifields and a
finite number of their derivatives.} or of a local functional is denoted by a
subscript, e.g. $\hat S=\sum_{k\geq 0} \hat S_k$.

In the context of the antifield
formalism, the existence of a Seiberg-Witten map translates into
the following four equivalent formulations.

\begin{itemize}
\item There exists a canonical
field-antifield transformation\footnote{
Only canonical transformation that
reduce to the identity at order $0$ in the deformation parameter are
considered here. Invertibility of these transformations in the space
of formal power series is then guaranteed.} $\hat\phi[\phi,\phi^*;\vartheta],
\hat\phi_*[\phi,\phi^*;\vartheta]$
such that
\bea
\hat
S[\hat\phi[\phi,\phi^*;\vartheta],\hat\phi_*[\phi,\phi^*;\vartheta];
\vartheta]=
S^{\rm eff}_0[\phi;\vartheta]+
\sum_{k\geq 1} S^{(0)}_k[\phi,\phi^*],\label{3.35}
\\ \Longleftrightarrow
\hat S[\hat\phi,\hat\phi_*;\vartheta]= S^{\rm
eff}_0[\phi[\hat\phi;\hat\phi^*;\vartheta];\vartheta]+ \sum_{k\geq 1}
S^{(0)}_k[\phi[\hat\phi,\hat\phi^*;\vartheta],\phi^*[\hat\phi,\hat\phi^*;
\vartheta]]
\label{3.36}, \eea
where $S^{\rm
eff}_0[\hat\phi;0]=S^{(0)}_0[\hat\phi;0]$.

\item There exists a generating functional of ``second type''
$F[\phi,\hat\phi^*;\defpar]$ in ghost number $-1$, with
\bea \hat\phi^A(x)=\frac{\delta^L
F}{\delta\hat\phi^*_A(x)},\ \phi^*_A(x) =\frac{\delta^L
F}{\delta\phi^A(x)},\label{transfo} \eea
such that
\bea
\hat S[\frac{\delta^L F}{\delta\hat\phi^*},\hat\phi_*;\defpar]=
S^{\rm eff}_0[\phi;\defpar]+\sum_{k\geq 1}
S^{(0)}_k[\phi,\frac{\delta^L F}{\delta \phi}],\label{3.38}
\eea
with initial condition $F=\int d^nx\
\hat\phi^*_A\phi^A+O(\vartheta)$.

\item The differential condition
\bea
\frac{\partial \hat S}{\partial \vartheta}
=\frac{\partial S^{\rm eff}_0}
{\partial \vartheta}+(\hat S,\hat \Xi)\label{3.37}
\eea
holds. The associated field-antifield redefinition satisfies the
differential equations
\bea
\frac{\partial \phi^A(x)}{\partial \vartheta}=(\phi^A(x),
\Xi(\vartheta)),\
\frac{\partial \phi^*_A(x)}{\partial \vartheta}=(\phi^*_A(x),
\Xi(\vartheta)), \label{eq:dsw}
\eea
where $\Xi (\vartheta)$ is the same function of the unhatted fields and
antifields than $\hat\Xi(\vartheta)$ is in terms of the hatted ones.
The formal solution to this differential equation
is given by \bea
\phi^A(x)=[P\exp\int_0^\vartheta d\vartheta^\prime
(\cdot,\hat\Xi(\vartheta^\prime))]\hat\phi^A(x),\
\phi^*_A(x)=[P\exp\int_0^\vartheta d\vartheta^\prime (\cdot,\hat
\Xi(\vartheta^\prime))]\hat\phi^A(x). \label{3.311} \eea

\item The deformed and undeformed theories are weakly gauge equivalent
  in the following sense.
In the case of an irreducible gauge theory, there
exists a simultaneous redefinition\footnote{A square bracket
  means a local dependence on the fields and their derivatives, while
  the round bracket means that this dependence is linear and
  homogeneous.} of the original gauge fields
$\hat\varphi^i=f^i[\varphi]$ and the parameters
$\hat\epsilon^\alpha=g^\alpha_\beta[\varphi](\epsilon^\beta)$ of the
irreducible generating set of nontrivial gauge transformations
$\hat R^i_\alpha[\hat\phi](\hat\epsilon^\alpha)$ such that
\begin{equation}
(\hat\delta_{\hat\epsilon} \hat\varphi^i\big)|_{\hat\varphi=f,\hat\epsilon=g}
\approx \delta_\epsilon f^i\,.
\end{equation}
Here $\approx$ means terms that vanish when the equations of motions
associated to $S^{\rm eff}_0[\varphi]\equiv \hat S_0[f[\varphi]]$ hold,
while
$\hat\delta_{\hat\epsilon}, \delta_\epsilon$ are respectively given by
\bea
  \hat\delta_{\hat\epsilon}&=\sum_{k=0}\partial_{\mu_1}\dots\partial_{\mu_k}
  \Big(\hat
  R^i_\alpha[\hat\varphi](\hat\epsilon^\alpha)\Big)\frac{\partial}{\partial
    (\partial_{\mu_1}\dots\partial_{\mu_k} \hat\varphi^i)},\\
  \delta_\epsilon&=\sum_{k=0}\partial_{\mu_1}\dots\partial_{\mu_k}\Big(
  R^i_\alpha[\varphi](\epsilon^\alpha)\Big)\frac{\partial}{\partial
    (\partial_{\mu_1}\dots\partial_{\mu_k} \varphi^i)}\,,
\eea
with
$R^i_\alpha[\varphi](\epsilon^\alpha)$ being the irreducible
generating set of nontrivial gauge transformations of the undeformed
theory.

\end{itemize}

Finally, the existence of a Seiberg-Witten map is controlled by the
BRST differential of the undeformed theory. In particular, if one can
show that in a relevant subspace, the representatives of the
cohomology of $s^{(0)}$ can be chosen to be antifield independent,
i.e., \bea (S^{(0)},C)=0\Longrightarrow
C=C^\prime_0+(S^{(0)},D),\label{ass} \eea
the Seiberg-Witten map are guaranteed to exist and can be
constructed as a succession of canonical transformations
\cite{Barnich:2001mc}.

\section{Remarks on noncommutative $U(N)$ Yang-Mills theory}

We assume the space-time manifold to be ${\bf R}^n$ with
coordinates $x^\mu\,,\mu=1,\dots,n$. The Weyl-Moyal star-product is
defined through \bea f*g
(x)=\exp\left(i{\wedge_{12}}\right)
f(x_1)g(x_2)|_{x_1=x_2=x},\ \ \ \wedge_{12}=
\frac{\vartheta}{2}{\theta^{\mu\nu}}
\partial_\mu^{x_1}\partial_\nu^{x_2}, \eea for a
real, constant, antisymmetric matrix $\theta^{\mu\nu}$. The
parameter $\vartheta$ has mass dimensions $-2$.
A minimal proper solution of the master
equation for noncommutative $U(N)$ Yang-Mills theory is given by:
\begin{equation}
\hat S=\int d^nx\ {\rm Tr}\ \left( -\frac{1}{4\kappa^2} \hat
F^{\mu\nu}*\hat F_{\mu\nu}+\hat A^{*\mu}* \hat D_\mu \hat
C+\frac{1}{2}\hat C^**[\hat C\stackrel{*}{,}\hat
C]\right)\,,
\label{eq:YM-maction}
\end{equation}
where fields, ghost fields, and their conjugated antifields
are $u(N)$ valued and $\rm Tr$ denotes ordinary matrix trace.
In particular, $\hat A^{*\mu}=\hat A^{*\mu}_Bg^{BA}T_A$, $\hat C^{*}=\hat
C^{*}_Bg^{BA}T_A$, with  $T_A$ being generators of Lie algebra $u(N)$
and $g_{AB}=\tr T_A T_B$ being an invariant metric on the algebra. We denote
by $[A\stackrel{*}{,}B]=A*B-(-)^{|A||B|}B*A$ the graded star commutator and
by $\{A\stackrel{*}{,}B\}=A*B+(-)^{|A||B|}B*A$ the graded star anticommutator.
Noncommutative Yang-Mills theory is a particular case of a consistent
deformation of standard Yang-Mills theory in the sense explained in the previous
section. The deformation parameter is $\vartheta$, the parameter of
"noncommutativity".

In the noncommutative Yang-Mills case, the differential condition
(\ref{3.37}) can be explicitly solved by
\begin{equation}\label{tdoea}
\frac{\partial S^{\rm eff}_0}{\partial
\vartheta}=\frac{1}{\kappa^2}\int d^nx\ {\rm Tr}\
\frac{i\theta^{\alpha\beta}}{2}\big(-\half\hat F^{ \mu\nu}\{\hat
F_{\alpha\mu}\stackrel{*}{,} \hat F_{\beta\nu}\}+\frac{1}{8}\{\hat
F^{\rho\sigma} \stackrel{*}{,}\hat F_{\rho\sigma}\}\hat
F_{\alpha\beta}\big),
\end{equation}
\begin{equation}\label{swgf}
\hat \Xi= \int d^nx\ {\rm Tr}\
\frac{i\theta^{\alpha\beta}}{2}\big( -\half\hat A^{* \mu}\{\hat
F_{\alpha\mu}+\d_\alpha\hat A_\mu\stackrel{*}{,}\hat
A_\beta\}+\half\hat C^{*}\{\hat A_\alpha\stackrel{*}{,}\d_\beta
\hat C\}\big),\label{Xi}
\end{equation}
The evolution equations $\partial \hat A/\partial\vartheta= -(\hat A,\hat \Xi)$
and $\partial \hat C/\partial\vartheta= -(\hat C,\hat \Xi)$
then reproduce the original differential equations of
\cite{Seiberg:1999vs}.

Linearity in antifields of the generating functional $\hat \Xi$
implies that the generating functional $F$ of second type can also
be chosen linear in antifields,
\bea
F=\int d^nx\ {\rm Tr}\
\left(\hat A^{*\mu} f_\mu+ \hat C^{*}\g\right)
\eea
where $f_\mu=f_\mu^A[A;\defpar]T_A$ and $\g=\g^A[A,C;\defpar]T_A$,
with $\g$ linear and homogeneous in the ghosts $C^A$ and their derivatives).
Linearity in antifields then implies that equation~(\ref{3.38})
reduces to

\bea
   \int d^nx \tr -\frac{1}{4\kappa^2}\hat F_{\mu\nu}\hat F^{\mu\nu}
  [f;\defpar]=S^{\rm eff}_0[A;\defpar]\\
  (\partial_\mu \g +[ f_\mu\stackrel{*}{,}\g])^A=\gamma f^A_\mu,\label{3.29}\\
 \frac{1}{2}\qcommut{\g}{\g}^A +\gamma \g^A=0\,,
\label{3.30}
\eea
where $\gamma$ is the gauge part of the BRST differential of the commutative
theory:
\bea
\gamma =\sum_{k=0}[\partial_{\rho_1}\dots  \partial_{\rho_k}
  (D_\mu C)^B\frac{\partial^L }{\partial (\partial_{\rho_1}\dots
\partial_{\rho_k}   A^B_{\mu})}-
\nonumber\\-
\frac{1}{2}\partial_{\rho_1}\dots
\partial_{\rho_k}(f^B_{DE}C^DC^E) \frac{\partial^L }{\partial
(\partial_{\rho_1}\dots  \partial_{\rho_k}C^B)}]\,. \eea
Note that for a generating functional $F$ of general
form, i.e., not necessarily linear in the antifields,
equations (\ref{3.29}) and (\ref{3.30}) would contain
equations of motion terms~\cite{Barnich:2001mc}.

Equation (\ref{3.29}) is the Seiberg-Witten equation (3.3) of
\cite{Seiberg:1999vs} under the form
$
\hat \delta_{\hat\lambda} \hat A=\delta_\lambda\hat A,$
with the identifications $\hat A_\mu\leftrightarrow f_\mu$, $\hat
\lambda\leftrightarrow \g$ and $\lambda\leftrightarrow C$.
When solving (\ref{3.29})-(\ref{3.30}),
it is useful to solve first the BRST version of the
integrability condition~(\ref{3.30}) before solving the Seiberg-Witten
equation (\ref{3.29}), because it contains as unknown functions only
the noncommutative gauge parameter $\g$ as a function of
$\theta^{\mu\nu}$, $C^A$ $A_\mu^A$ and their derivatives.

\pagebreak

{\bf Remarks}:

\begin{enumerate}

\item  The existence of the Seiberg-Witten map can be infered a priori from the
knowledge of the local BRST cohomology of commutative Yang-Mills theory
\cite{BBH}. The point is that in the
infinitesimal noncommutative deformation, only differentiated ghosts appear in
the antifield dependent terms while the local cohomology in ghost number $0$ of
the BRST differential $s^{(0)}$ of commutative Yang-Mills theory can be shown to
depend only on undifferentiated antifields and undifferentiated ghosts. This
implies the existence of the Seiberg-Witten generating functional to first order
in the deformation parameter $\defpar$. Furthermore, one can show that this
reasoning can be iterated, which allows to prove the existence of the
Seiberg-Witten map along the lines of section 2.5 of \cite{Barnich:2001mc}.

\item Besides the expansion in the deformation parameter $\defpar$, it is often
useful to consider an expansion in the homogeneity of the fields. The BRST
differential of the noncommutative theory then reduces to lowest order to the
BRST differential $s^{[0]}$ of $N$ free commutative abelian fields. In both
expansions, the contracting homotopy that allows to invert $s^{(0)}$
respectively $s^{[0]}$ in the relevant subspace can be explicitly constructed by
a simple change of generators that consists essentially in replacing the
derivatives of the gauge potentiels by the symmetrized derivatives
of the gauge potentials and the symmetrized covariant derivatives of the field
strenghts. This last symmetrization includes the first index of the field
strength in order to get rid of the redundancies due to the Bianchi
identities (see e.g. \cite{hom,BBH}
for details). This explicit form of the contracting homotopy operator allows to
construct the generating functional $\hat\Xi$ of equation (\ref{Xi}) and to
solve the gauge equivalence conditions recursively (see also
\cite{Brace:2001fj}).

\item A natural question to ask is whether the whole noncommutative deformation
is trivial in the sense that it can be undone by a field redefinition. This
question can also be addressed using local BRST cohomology. Indeed, in the
Yang-Mills case, it turns out that the infinitesimal deformation of the action
corresponds to a non trivial BRST cohomology class in ghost number $0$ implying
the non triviality of the noncommutative deformation. In noncommutative
Chern-Simons theory however, results on the local BRST cohomology of standard
Chern-Simons theory imply that the whole noncommutative deformation is trivial.
This has been shown directly in \cite{Grandi:2000av}.

\item After gauge fixing, which corresponds merely to another canonical
transformation in the antifield formalism, the
appropriate formulation to control the Seiberg-Witten map during perturbative
renormalization is the functional differential equation (\ref{3.37}), which
should be promoted to an analogous equation for the generating functional for
1PI Green's function, in the same way than the master equation (\ref{meq})
gets promoted to the Zinn-Justin equation and controls the gauge invariance.
Since it is well known how the second term on the right hand side
of (\ref{3.37}) renormalizes (see e.g. \cite{Voronov:1982ph}), the question
reduces to the renormalization of the higher dimensional
operator ${\partial S^{\rm eff}_0}/{\partial
\vartheta}$ given in (\ref{tdoea}).

\end{enumerate}

\section*{Acknowledgments}

The authors want to thank M.~Henneaux for useful discussions. G.B. is "Chercheur
qualifi\'e du Fonds National de la Recherche Scientifique". This research has
been partially supported by the ``Actions de Recherche Concert{\'e}es" of the
``Direction de la Recherche Scientifique - Communaut{\'e} Fran{\c c}aise de
Belgique", by IISN - Belgium (convention 4.4505.86),  by
Proyectos FONDECYT 1970151 and 7960001 (Chile),
by the European Commission RTN programme HPRN-CT-00131,
in which the authors are associated to K. U. Leuven and
by INTAS grant 00-00262. The work of M. G. is partially supported by
the RFBR grant 01-01-00906.




\providecommand{\href}[2]{#2}\begingroup\raggedright\endgroup

\end{document}